\documentclass[fleqn,usenatbib]{mnras}

\usepackage{newtxtext,newtxmath}

\usepackage[T1]{fontenc}

\DeclareRobustCommand{\VAN}[3]{#2}
\let\VANthebibliography\thebibliography
\def\thebibliography{\DeclareRobustCommand{\VAN}[3]{##3}\VANthebibliography}

\usepackage{graphicx}	
\usepackage{amsmath}	
\usepackage{hyperref}
\usepackage{orcidlink}
\usepackage{xspace}
\usepackage{textcase}
\usepackage{xstring}
\usepackage{xcolor}
\usepackage[mathlines]{lineno}

\title[Kozai Lidov Cycles = Simple Pendulum]{Kozai Lidov Cycles = Simple Pendulum}

\author[R. Basha et al.]{
Roi D. Basha\orcidlink{0009-0009-3571-6192}\thanks{E-mail: rdbasha0@gmail.com (RDB)},
Ygal Y. Klein\orcidlink{0009-0004-1914-5821}
and Boaz Katz\orcidlink{0000-0003-0584-2920}
\\
Dept. of Particle Phys. \& Astrophys., Weizmann Institute of Science,
 Rehovot 76100, Israel
}

\date{Accepted XXX. Received YYY; in original form ZZZ}

\pubyear{\the\year{}}

\begin{document}
\label{firstpage}
\pagerange{\pageref{firstpage}--\pageref{lastpage}}
\maketitle

\begin{abstract}
The quadrupole Kozai mechanism, which describes the hierarchical three-body problem in the leading order, is shown to be equivalent to a simple pendulum where the change in the eccentricity squared equals the height of the pendulum from its lowest point: $e_{\text{max}}^2-e^2=h=l\left(1-\cos{\theta}\right)$. In particular, this results in useful expressions for the KLC period, and the maximal and minimal eccentricities in terms of orbital constants. We derive the equivalence using the vector coordinates $\boldsymbol{\alpha}=\textbf{j}+\textbf{e}, \boldsymbol{\beta}=\textbf{j}-\textbf{e}$ for the inner Keplerian orbit, where $\textbf{j}$ is the normalized specific angular momentum, and $\textbf{e}$ is the eccentricity vector. The equations of motion for $\boldsymbol{\alpha}$ and $\boldsymbol{\beta}$ simplify to $\dot{\boldsymbol{\alpha}}=2\partial_{\boldsymbol{\alpha}} \phi \times \boldsymbol{\alpha}$ and $\dot{\boldsymbol{\beta}}=2\partial_{\boldsymbol{\beta}} \phi \times \boldsymbol{\beta}$, where $\phi$ is the normalized averaged interaction potential, and are symmetric to replacing $\boldsymbol{\alpha}$ and $\boldsymbol{\beta}$ for the KLC quadratic potential. Their constraints simplify to $\boldsymbol{\alpha}^2=\boldsymbol{\beta}^2=1$, and they are distributed uniformly and independently on the unit sphere for a uniform distribution in phase space (with a fixed energy). 
\end{abstract}

\begin{keywords}
gravitation-celestial mechanics-planets and satellites: dynamical evolution and stability-stars: multiple: close
\end{keywords}





\section{The map between Kozai-Lidov Cycles (KLCs) and a simple pendulum }


We consider the hierarchical three-body problem, where a test particle orbits a primary object of mass $m$ with semi-major axis $a$ and eccentricity $e$, and is perturbed by a distant third object of mass $m_{\text{per}}$, moving along an outer orbit characterized by semi-major axis $a_{\text{per}}$ and eccentricity $e_{\text{per}}$. We focus on the limit $a/a_{\text{per}}\ll1$ where the perturbing potential can be expanded to quadrupole order and the equations of motion can be averaged over the inner and outer orbits, which has been solved analytically and results in periodic oscillations of the inner eccentricity and inclination (Kozai-Lidov Cycles, \cite{kozai1962,lidov1962}). The z-axis is chosen in the direction of (constant) outer angular momentum and the $x$-axis is chosen in the (constant) direction of the outer orbit's eccentricity vector $\mathbf{e_{\text{per}}}$. 
The inner orbit is parametrized by the eccentricity vector $\mathbf{e}$ and normalized specific angular momentum $\mathbf{j}=\mathbf{J}/\sqrt{Gma}$ that satisfy $\mathbf{j}\cdot\mathbf{e}=0$ and $j^2+e^2=1$.

Expanding the perturbing potential to the leading (quadrupole) order term in the parameter $a/a_{\text{per}}\ll1$, and averaging over the inner and outer orbits results in the potential: $\Phi_{{\text{per}}}=\Phi_{0}\phi_{\text{quad}}$ where
 $\phi_{\text{quad}}=\frac{3}{4}\left(\frac{5}{2}e_{z}^{2}-e^2-\frac{1}{2}j_{z}^{2}+\frac{1}{6}\right),\label{eq:phi_quad}$
$j_z=\sqrt{1-e^2}\cos i$ and $e_z=e\sin i\sin\omega$, are the $z$ components of $\mathbf{j}$ and $\mathbf{e}$ respectively, $i$ is the inclination, $\omega$ is the argument of pericenter and
$\Phi_{0}=Gm_{\text{per}}a^{2}a_{\text{per}}^{-3}(1-e_{\text{per}}^{2})^{-3/2}$ is a constant normalization factor.
For plots and derivatives, we use the normalized time $\tau\equiv t/t_\text{sec}$, where $t_\text{sec}=(\sqrt{Gma}/\Phi_0)$ is the secular time.

KLCs can be parametrized by the following two constants of motion: $j_z$ (due to the axisymmetry of the perturbing potential) and 
\begin{align}
C_{K}
=e^{2}-\frac{5}{2}e^2_z=e^2\left(1-\frac{5}{2}\sin^2 i \sin^2 \omega\right),\label{eq:CK}
\end{align}
which is a linear combination of the constants $j_z^2$ and $\phi_{\rm quad}$.

In section \S\ref{sec:demonstration} we show that the equations that govern the KLCs in the quadrupole approximation are equivalent to that of a mechanical pendulum:
\begin{equation}
\ddot{\theta}=-\omega_0^2 \sin{\theta},\label{eq:pendulum}
\end{equation}
where $\theta$ is the pendulum angle and $\omega_0$ is a constant (equal to the angular frequency at small oscillations). 
The correspondence is given by:
\begin{align}
\dot{\theta}&=\frac32\sqrt{15}~e_z=\frac32\sqrt{15}~e\sin i \sin \omega,\label{eq:theta_dot}\\
\omega_0^2&=\frac{9}{8}\sqrt{(3-5j_z^2-2C_K)^2 + 24C_K}\label{eq:omega0}\\
E&=1+\frac{9}{8\omega_0^2}\left(3-5j_z^2-8C_K\right),\label{eq:pendulum_energy}
\end{align}
where
\begin{align}
  E=\frac{1}{2\omega_0^2}\dot{\theta}^2+1-\cos{\theta}\label{eq:pendulum_energy_def} 
\end{align}
is the dimensionless constant of motion of the pendulum that corresponds to the energy. 

\begin{figure}
\includegraphics[width=\columnwidth]{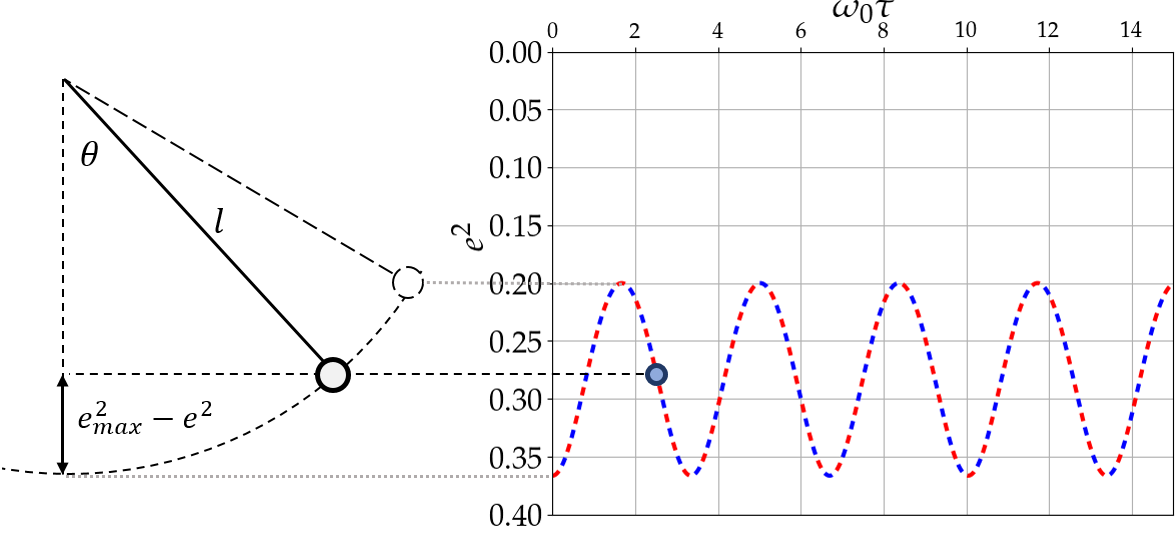}
\caption{Illustration of the equivalence between a simple pendulum and KLC. The plot shows $e^2$ as a function of (normalized) time calculated in two ways which exactly agree. Blue: A numerical integration of the KLC equations \eqref{eq:jedot}. Red: A numerical integration of the pendulum equation \eqref{eq:pendulum} mapped to $e^2$ using equation \eqref{eq:e^2}. The KLC parameters are $j_z=0.72$ and $C_K=0.2$ with the corresponding pendulum parameters $\omega_0\approx 1.6$ (equation \ref{eq:omega0}) and $E\approx0.46$ (equation \ref{eq:pendulum_energy}). Since $C_K>0$ (and so $E<2$), the argument of periapsis rotates, and the pendulum librates, schematically shown in the diagram on the left. Note that the \textit{y}-axis is reversed.}
\label{fig:PendulumLib}
\end{figure}

\begin{figure}
\includegraphics[width=\columnwidth]{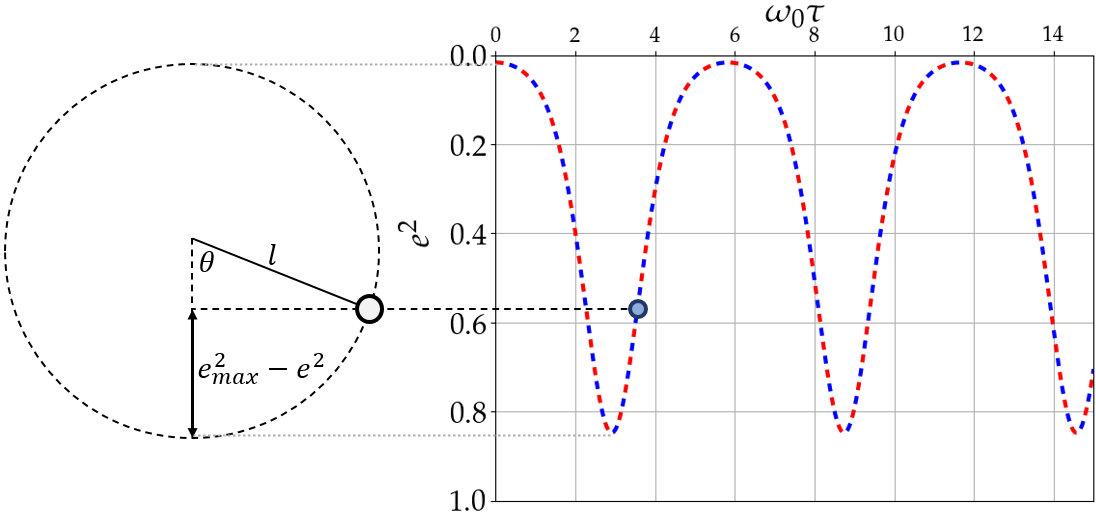}
\caption{Same as Fig. \ref{fig:PendulumLib} but for the parameters $j_z=0.3$ and $C_K=-0.021$ (corresponding to $\omega_0\approx 1.7$ and $E\approx 2.1$) where the argument of periapsis librates, and the pendulum rotates.}
\label{fig:PendulumRot}
\end{figure}

A simple relation between the properties of the pendulum and the eccentricity can be obtained from equations \eqref{eq:CK},\eqref{eq:theta_dot}, and \eqref{eq:pendulum_energy_def} by choosing the (equivalent) pendulum \textit{"length"} to be:
\begin{equation}
l=\frac{4}{27}\omega_0^2,\label{eq:pendulum_length}
\end{equation}
resulting in:
\begin{equation}
e^2=lE+C_K-l(1-\cos\theta)=e_{\max}^2-h\label{eq:e^2},
\end{equation}
where $h=l(1-\cos\theta)$ is the height of the pendulum from its lowest point and the maximal eccentricity is obtained at the lowest point, $\theta=0$.
A demonstration of the equivalence is shown in Figs. \ref{fig:PendulumLib} and \ref{fig:PendulumRot} for parameters corresponding to libration and rotation of the pendulum respectively. In the plots, direct numerical integrations of the pendulum equation (in red, equations \ref{eq:pendulum}-\ref{eq:e^2}) are over-plotted on direct integrations of the KLC equations (in blue, equations \ref{eq:jedot} in section \ref{sec:coordinates} below) aligning perfectly. 

Several consequences of the correspondence are next elaborated. First, note that equation \eqref{eq:theta_dot} implies that when the pendulum rotates (i.e. $\dot\theta\neq 0$) the argument of pericenter librates ($\sin\omega \neq 0$) and vice-versa. Therefore $E > 2$ ($E < 2$)  corresponds to $C_K<0$ ($C_K>0$). Second, the minimal and maximal eccentricities can be readily obtained by the pendulum constants as follows. Maximal eccentricity occurs at $\theta=0$ implying (see equation \ref{eq:e^2}): 
\begin{align}\label{eq:e_max^2}
e_{\max}^2=lE+C_K.
\end{align}
The minimal eccentricity is obtained when $h$ is maximal, and its expression depends on the mode of motion. For pendulum libration, the minimal eccentricity $e_\text{min}$ ($h_\text{max}$) occurs when the pendulum stops, $\dot{\theta}\propto e_z=0$, and equation \eqref{eq:CK} implies $C_K=e_\text{min}^2$. For pendulum rotation, minimal eccentricity occurs in the vertical position $h=2l$ and equation \eqref{eq:e^2} implies $e^2=l(E-2)+C_K$. Therefore:
\begin{equation}
e_\text{min}^2= \begin{cases} \label{eq:e_min^2}
          C_K, & C_K>0 \\
          l\left(E-2\right)+C_K, & C_K<0 
       \end{cases}
\end{equation}
The oscillation period of $e$ equals the period of the pendulum's height, which is half the period of the pendulum angle and therefore,
\begin{equation}
T_{\text{KL}}=\frac12T_{\rm Pen}t_{\rm sec}=\frac{2}{\omega_0}K\left( \sqrt{\frac{E}{2}}\right)t_\text{sec},\label{eq:period}
\end{equation}
where $K(s)$ denotes the complete elliptic integral of the first kind.
Note that when the pendulum rotates, $E>2$, $K(\sqrt{E/2})$ is complex, and the period equals the real part of the result.

The expressions for $e_{\max}$ and $e_{\min}$ here (equations \ref{eq:e_max^2}, \ref{eq:e_min^2}) agree with equations 28-31 in \cite{antognini2015}. The expression for the period here (equation \ref{eq:period}) agrees with numerical evaluations of the integral expression in equation 27 in \cite{antognini2015}. 

One of the striking features of Kozai oscillations is the difference in behavior at small $e$ for inclinations below or above the critical inclination $i_{\rm crit}=39.2^\circ$ \citep[$j_{z,\rm crit}^2=\cos^2(i_\text{crit})=3/5$,][]{kozai1962,lidov1962}. This can be explained within the pendulum model as follows. 
For $e\rightarrow0$ ($C_K\rightarrow0$), equations \eqref{eq:theta_dot}-\eqref{eq:pendulum_energy} imply $\dot\theta\rightarrow0$ and $E\rightarrow 1+\text{sign}(3-5j_z^2)$. For $i<i_{\rm crit}$, $j_z^2>3/5$ and $E\rightarrow0$ (slow pendulum near bottom) implying small oscillations in eccentricity, while for $i>i_{\rm crit}$, $j_z^2<3/5$ and $E\rightarrow 2$ (slow pendulum near top) implying large oscillations in eccentricity.
\section{Useful Keplerian coordinates}\label{sec:coordinates}
The poisson brackets of vectors $\mathbf{j},\mathbf{e}$ admit an $\mathfrak{so(\text{4})}$ algebra \citep[e.g.][]{tremaine2009}: 
\begin{equation}
\begin{aligned}
\{j_i,j_j\}&=\frac{1}{\sqrt{GMa}}\epsilon_{ijk}j_k, \\
\{j_i,e_j\}&=\frac{1}{\sqrt{GMa}}\epsilon_{ijk}e_k, \\
\{e_i,e_j\}&=\frac{1}{\sqrt{GMa}}\epsilon_{ijk}j_k.
\end{aligned}
\end{equation}

The algebra can be split into two copies of $\mathfrak{su(\text{2})}$, by defining the vectors (e.g. chapter 2 in \citet{Ryder_1996} for the general case, and \citet{Weinberg2011} for the Keplerian problem):
\begin{equation}\label{eq:alpha_beta_def}
\begin{aligned}
    \boldsymbol{\alpha} &= \mathbf{j}+\mathbf{e},\\
    \boldsymbol{\beta}  &= \mathbf{j}-\mathbf{e},
\end{aligned}
\end{equation}
resulting in the simpler relations:
\begin{equation}
\begin{aligned}
\{\alpha_i,\alpha_j\}&=\frac{2}{\sqrt{GMa}}\epsilon_{ijk}\alpha_k,\\
\{\beta_i,\beta_j\}&=\frac{2}{\sqrt{GMa}}\epsilon_{ijk}\beta_k,\\
\{\alpha_i,\beta_j\}&=0.
\end{aligned}
\end{equation}
Using $\dot{f}=\{f,\phi\}=\{f,\eta_l\}\frac{\partial \phi}{\partial \eta_l}$ for any function $f$ and phase space coordinates $\eta_l$, the corresponding equations of motion are (e.g \cite{milankovitch1939,tremaine2009})
\begin{equation}\label{eq:jedot}
\begin{aligned}
\dot{\mathbf{j}}&=-\mathbf{j}\times \frac{\partial \phi}{\partial \mathbf{j}}   - \mathbf{e}\times\frac{\partial \phi} {\partial \mathbf{e}},\\
\dot{\mathbf{e}}&=-\mathbf{j}\times \frac{\partial \phi}{\partial \boldsymbol{e}} - \mathbf{e}\times \frac{\partial \phi}{\partial \mathbf{j}},
\end{aligned}
\end{equation}
and
\begin{equation}\label{eq:ABdot}
\begin{aligned}
\dot{\boldsymbol{\alpha}}&=2\frac{\partial \phi}{\partial \boldsymbol{\alpha}} \times \boldsymbol{\alpha},\\
\dot{\boldsymbol{\beta}}&=2\frac{\partial \phi}{\partial \boldsymbol{\beta}} \times \boldsymbol{\beta}. 
\end{aligned}
\end{equation}
Finally, the algebraic relations $j^2+e^2=1$ and $\boldsymbol{e}\cdot \boldsymbol{j}=0$ translate to 
\begin{equation}\label{eq:alpha_beta_constraints}
\alpha^2=\beta^2=1.
\end{equation}

The coordinates $\boldsymbol{\alpha}$ and $\boldsymbol{\beta}$ have several advantages over $\mathbf{j}$ and $\mathbf{e}$. First, the equations of motion \eqref{eq:ABdot} are simpler than equations \eqref{eq:jedot} (note that their equivalence can be deduced directly from the definition \eqref{eq:alpha_beta_def}). Second, the constraints on $\boldsymbol{\alpha}$ and $\boldsymbol{\beta}$ (equations \ref{eq:alpha_beta_constraints}) are simpler than those of $\mathbf{j}$ and $\mathbf{e}$. In particular, it is useful that they are separate. Third, the fact that $\alpha^2,\beta^2,\phi$ are constants can be more easily seen from equations \eqref{eq:ABdot} than the fact that $j^2+e^2, \mathbf{j}\cdot \mathbf{e}, \phi$ are constants from equations \eqref{eq:jedot}. Finally, using
\begin{equation}\label{eq:relation_e2_alpha_beta}
 \boldsymbol{\alpha} \cdot \boldsymbol{\beta}=1-2e^2,
 \end{equation}
a uniform distribution in phase space corresponds to independent isotropic distributions of $\boldsymbol{\alpha}$ and $\boldsymbol{\beta}$ on the unit spheres, which is simpler than the corresponding distribution of $\mathbf{j}$ and $\mathbf{e}$.

\section{Demonstration of the equivalence between KLC and a simple pendulum}\label{sec:demonstration}

In terms of $\boldsymbol{\alpha},\boldsymbol{\beta}$ the normalized quadrupole potential can be expressed as:
\begin{align}\label{eq:alpha_beta_phi}
\phi_{\rm quad}=\frac38\left(\boldsymbol{\alpha}\cdot\boldsymbol{\beta}+(\alpha_z-\beta_z)^2-\alpha_z\beta_z\right)-\frac14,
\end{align}
resulting with the equations of motion:
\begin{equation}\label{eq:alpha_beta_eom}
\begin{aligned}
\dot{\boldsymbol{\alpha}}&=\frac34\left(\boldsymbol{\beta}+(2\alpha_z-3\beta_z)\hat z\right)\times \boldsymbol{\alpha},\\
\dot{\boldsymbol{\beta}}&=\frac34\left(\boldsymbol{\alpha}+(2\beta_z-3\alpha_z)\hat z\right)\times \boldsymbol{\beta}.
\end{aligned}
\end{equation}
Note that equations \eqref{eq:alpha_beta_phi} and \eqref{eq:alpha_beta_eom} are symmetric to a swap of $\boldsymbol{\alpha}$ and $\boldsymbol{\beta}$ providing an additional advantage to these coordinates in this case. This symmetry is a result of the symmetry $\phi_{\rm quad}(-\mathbf{e})=\phi_{\rm quad}(\mathbf{e})$ in the quadrupole potential and the symmetric form of the general equations \eqref{eq:ABdot}.

By defining:
\begin{equation}\label{eq:csv_def}
\begin{aligned}
&v=\frac34\sqrt{15}~(\alpha_z-\beta_z),\\ 
&s=\frac{9}{8}\sqrt{15}~(\boldsymbol{\alpha}\times\boldsymbol{\beta})_z, \\ 
&c=\frac{9}{8}(1+5\alpha_z\beta_z-4\boldsymbol{\alpha} \cdot \boldsymbol{\beta}),  
\end{aligned}
\end{equation}
the following closed set of equations are obtained: 
\begin{equation}\label{eq:vsc}
\begin{aligned}
&\dot{v}=-s,\\
&\dot{s}=vc,\\
&\dot{c}=-vs,  
\end{aligned}
\end{equation}
which are equivalent to the equations of a pendulum.
Indeed, equations \eqref{eq:vsc} are equivalent to the pendulum equation \eqref{eq:pendulum} by the substitution: 
\begin{equation}\label{eq:cs_decomposition}
\begin{aligned}
&\omega_0^2=\sqrt{c^2+s^2}=\rm const.,\\
&v= \dot{\theta},\\
&s= \omega_0^2 \sin{\theta},\\
&c= \omega_0^2 \cos{\theta},
\end{aligned}
\end{equation}
with the additional constant of motion (normalized pendulum energy):
\begin{align}\label{eq:E_def_vsc}
&E= \frac{1}{\omega_0^2}\left(\frac{v^2}{2}-c\right)+1=\rm const.
\end{align}
Equations \eqref{eq:theta_dot}-\eqref{eq:pendulum_energy} can be obtained from equations \eqref{eq:CK},\eqref{eq:alpha_beta_def},\eqref{eq:csv_def}, \eqref{eq:cs_decomposition} and \eqref{eq:E_def_vsc} by direct substitution. 
We note that while we originally derived the equivalence using $\boldsymbol{\alpha}$ and $\boldsymbol{\beta}$ as presented above it can also be derived directly from the equations for $\mathbf{j}$ and $\mathbf{e}$. 

\section{Conclusions}
It was recently shown that two distinct problems involving the long-term evolution of KLCs under additional perturbations can be mapped to simple pendulum dynamics in the limit of high eccentricity \citep[i.e. low $\left|j_z\right|$,][]{Klein2024_pendulum,Klein2024_CDA,klein2025}. The perturbations include the next order term in the multi-pole expansion \citep[octupole, e.g.][]{naoz2011,katz2011,naoz2016} with or without time-dependent corrections \citep[e.g. ][]{luo2016,Will2021,tremaine2023} as well as a constant precession of the outer orbit \citep[][]{hamers2017,petrovich2017,Klein2023,Klein2024}.   
The equivalence shown here is complementary and differs in two key respects. First, these previous results are approximate and relate to the evolution of the KLC parameters (in particular $j_z$) over time scales much longer than a single KLC while the equivalence here is exact and deals with the evolution of the Keplerian parameters within one KLC. Secondly, these previous results are obtained in the limit $\left|j_z\right|\ll 1$, while the equivalence shown here is for arbitrary $j_z$. 
In the limit $j_z=0$, \citet{klein2025} have shown (appendix A therein) that the unperturbed KLC system itself becomes equivalent to a simple pendulum. This mapping is reproduced from equations \eqref{eq:pendulum}-\eqref{eq:pendulum_energy_def} by setting $j_z=0$ and $1-\cos{\theta'}=(2/E)(1-\cos{\theta})$, where $\theta'$ is the pendulum angle in \cite{klein2025}. Note that for any pendulum such a mapping results in an equivalent pendulum equation for the new angle $\theta'$ with $\omega_0'^2=\omega_0^2E/2$, where libration of the one corresponds to rotation of the other.

Finally, there are two natural extensions to the validity of the equivalence presented here. First, the results remain valid for non-test particle systems providing that the inner angular momentum is much smaller than the outer (as do KLCs in general) with the mass $m$ taken to be the sum of the inner binary components. Second, the equivalence can be extended straightforwardly to any axisymmetric quadrupole potential, such as that induced by a galactic potential on wide orbits \citep[e.g.,][]{Heisler1986}.



\bibliographystyle{mnras}
\bibliography{Kozai_pendulum} 





\bsp	
\label{lastpage}
\end{document}